\documentstyle[mprocl]{article}

\def\be{\begin{equation}}
\def\ee{\end{equation}}
\def\bea{\begin{eqnarray}}
\def\eea{\end{eqnarray}}

\def \a {\alpha}

\def \ga {\gamma}

\def \part {\partial}

\def \um {{1\over 2}}

\def\IR{{\hbox{{\rm I}\kern-.2em\hbox{\rm R}}}}
\def\IT{{\hbox{{\rm I}\kern-.2em\hbox{\rm T}}}}
\def\I{{\hbox{{\rm I}\kern-.2em\hbox{\rm I}}}}
\def\IH{{\hbox{{\rm I}\kern-.2em\hbox{\rm H}}}}
\def\IC{{\hbox{\kern-.2em\hbox{\bf C}}}}
\def\IZ{{\hbox{{\rm Z}\kern-.4em\hbox{\rm Z}}}}

\begin{document}
gr-qc/9710131

\title{CONSTANTS OF THE MOTION AND THE QUANTUM MODULAR GROUP IN (2+1)- 
DIMENSIONAL GRAVITY}

\author{ V. MONCRIEF}

\address{ Department of Physics and Department of Mathematics,\\
Yale University, 217 Prospect Street,\\
New Haven, Conn. 06511, USA}

\author{ J.E.NELSON}

\address{ Dipartimento di Fisica Teorica dell'Universit\`a di Torino,\\
via Pietro Giuria 1, 10125 Torino, Italy}


\maketitle\abstracts{
Constants of motion are calculated for 2+1 dimensional gravity with
topology $\IR \times T^2$ and negative cosmological constant. Certain 
linear combinations of them satisfy the 
anti - de Sitter algebra $\hbox{so}(2,2)$ in either ADM or holonomy 
variables. Quantisation is straightforward in terms of the holonomy 
parameters, and the modular group is generated 
by these conserved quantities. On inclusion of the Hamiltonian three 
new global constants are derived and the 
quantum algebra extends to the conformal algebra $\hbox{so}(2,3)$.}
\section{Hamiltonian Dynamics}\label{subsec:prod}
It is known \cite{mn,cn} that 
(2+1)-dimensional gravity, with or without a cosmological constant 
$\Lambda$, has (at least)
two equivalent descriptions. For topology $\IR \times T^2$ and $\Lambda 
< 0$ the $\tau$ = extrinsic curvature development of the ADM canonical 
variables $q^1,q^2,p_1,p_2$ is generated by an effective Hamiltonian 
which is just the spatial volume \cite{vm1,hos}
\be
H = \int_{T^2} d^2x \sqrt{^{(2)}g} = {1 \over  
\sqrt{\tau^2 - 4\Lambda}} ~\bar{H}, \qquad 
\bar{H} = \sqrt{p_1^2 + e^{-2q^1}p_2^2}.
\label{11}
\ee
An alternative description \cite{nr1} is in terms of global, 
time-independent 
parameters $r_{1,2}^{\pm}$ which characterise the traces of holonomies 
(Wilson loops) \cite{cn} and satisfy 
\be\{r_1^\pm,r_2^\pm\}=\mp {1\over\a}, \qquad \{r^+,r^-\}= 0, \qquad
and~~ \a = {1 \over \sqrt{-  \Lambda}} > 0.
\label{12}
\ee
In Eq.~\ref{12} the subscripts 1, 2 refer to two intersecting 
paths $\ga_1, \ga_2$ on $T^2$ with intersection number $+1$, and the 
$\pm$ refer to the two copies of $\hbox{SL(2,$\IR$)}$
in the decomposition of the spinor group of 
$\hbox{SO}(2,2)$ 
as a tensor product $\hbox{SL(2,$\IR$)}\otimes\hbox{SL(2,$\IR$)}$ 
\cite{nr1}.

The four real parameters $r_1^{\pm}, r_2^{\pm}$ of Eq.~\ref{12} are 
arbitrary, but are related, 
through a time-dependent canonical transformation, to the 
components of the complex moduli $m = m_1 +im_2$ and their momenta 
$\pi=\pi^1 +i \pi^2$ as follows \cite{cn}
\bea
m & = &\left(r_1^-e^{it/\a} + r_1^+e^{-{it/\a}}\right) 
 \left(r_2^-e^{it/\a} + r_2^+e^{-{it/\a}}\right)^{\lower2pt%
 \hbox{$\scriptstyle -1$}}, \nonumber \\
\pi & = & -{i\a\over 2\sin{2t\over \a}}\left(r_2^+e^{it/\a} 
 + r_2^-e^{-{it/\a}}\right)^{\lower2pt%
 \hbox{$\scriptstyle 2$}}, 
\label{13}
\eea
where
\be
m_1 = q^2, m_2 = e ^{-q^1}, \pi^1 = p_2, \pi^2 = -p_1 
e^{q^1} \qquad and~~ \tau = -{2 \over {\a}}\cot {2t \over  {\a}}, 
\label{14}
\ee
with $\tau$ monotonic in the range $t~ \epsilon~ (0, {\pi \a \over 2})$, 
and from Eqs.~\ref{13}--\ref{14} the Hamiltonian 
(Eq.~\ref{11}) is
\be
H = {\a \over {2\sqrt{\tau^2 - 4\Lambda}}}(r_1^-r_2^+ 
- r_1^+r_2^-),
\quad \bar H = {\a \over 2} ( r_1^- r_2^+ 
-  r_1^+ r_2^-).
\label{15}
\ee
Since the $r_1^{\pm}, r_2^{\pm}$ are arbitrary  the moduli and momenta 
of Eq.~\ref{13} can have arbitrary initial data $m(t_0), p(t_0)$ at some 
initial time $t_0$. 
\section{Constants of the motion}
\subsection{ADM Variables}
Absolutely conserved quantities are obtained \cite{mn}, as in 
\cite{vm2,mart}, 
from the traces of the  $\hbox{SL(2,$\IR$)}$
holonomies along the paths $\ga_1, \ga_2, \ga_1\cdot\ga_2$ respectively, 
\be
C_1^{\pm} =C_1 \pm 2\sqrt{-\Lambda} C_4, \quad
C_2^{\pm} =C_2 \mp 2\sqrt{-\Lambda} C_5, \quad
C_3^{\pm} =C_3 \pm \sqrt{-\Lambda} C_6,
\label{21}
\ee
where
\begin{eqnarray}
&C_1 &= {1 \over 2} ~e^{-q^1} \tau \left\{ (\sqrt{1 - {4\Lambda \over  
\tau^2}} \bar{H} - p_1)(1 + (q^2)^2 e^{2q^1}) - 2(q^2 p_2 - p_1)  
\right\} , \nonumber \\
&C_2 &= {1 \over 2} ~e^{q^1} \tau\left\{\sqrt{1 - {4\Lambda \over  
\tau^2}} \bar{H} - p_1 \right\} , \nonumber \\
&C_3 &= {1 \over 2} ~e^{q^1} \tau\left\{ q^2 (\sqrt{1 - {4\Lambda  
\over \tau^2}} \bar{H} - p_1) - p_2  ~e^{-2q^1} \right\} , \nonumber 
\end{eqnarray}
\be
C_4 = {1 \over 2} \left\{ p_2 ~e^{- 2q^1} + 2 q^2 p_1 - p_2 (q^2)^2  
\right\} , \quad
C_5 = {1 \over 2} ~p_2, \quad
C_6 = p_1 - q^2 p_2.
\label{22}
\ee
Quantisation of these constants and the Hamiltonian (Eq.~\ref{11}) 
in terms of these ADM variables has been discussed in \cite{car,puz}.
\subsection{Holonomy parameters}
In terms of the time independent parameters $r_{1,2}^{\pm}$ 
the $C_1^{\pm}  - C_3^{\pm}$ are, from Eqs.~\ref{13}--\ref{14}, 
\ref{21}--\ref{22},
\be
C_1^{\pm} = (r_1^{\mp})^2, \quad C_2^{\pm} = (r_2^{\mp})^2, \quad
C_3^{\pm} = r_1^{\mp}r_2^{\mp},
\label{23}
\ee
and they are evidently time independent. Quantisation is straightforward 
in terms of these parameters. With the commutators (the quantisation of 
Eq.~\ref{12})
\be
[{\hat r}_1^{\pm}, {\hat r}_2^{\pm}] =  \mp {i {\hbar} \over \a},
\qquad [{\hat r}^+,{\hat r}^-] = 0,
\label{24}
\ee
the combinations
\be
{\hat j}_0^{\pm} = \mp {\a \over 2} ({({\hat r}_1^{\pm})}^2 
+ {({\hat r}_2^{\pm})}^2), \quad
{\hat j}_1^{\pm} = \pm {\a \over 2} ({({\hat r}_1^{\pm})}^2 
- {({\hat r}_2^{\pm})}^2), \quad
{\hat j}_2^{\pm} = \pm {\a\over 2}
({\hat r}_1^{\pm}{\hat r}_2^{\pm} + {\hat r}_2^{\pm}{\hat r}_1^{\pm}),   
\label{25}
\ee
satisfy the two $(\pm)$ Lie algebras of $\hbox{so}(1,2) \approx 
\hbox{sl(2,$\IR$)}$.
\be
[{\hat j}_a^{\pm},{\hat j}_b^{\pm}] = 2i {\hbar} \epsilon_{abc}
{{\hat j}^c}{}^{\pm},  \quad [{{\hat j}_a}^+,{{\hat j}_b}^-] = 0,  
\label{26}
\ee
where the $j_a^+$ depend only on the $r^+$'s and the $j_a^-$ 
only on the $r^-$'s.

The generators ${\hat j}_a^{\pm}$ and $\hat {\bar H}$ are not all 
independent. There are 3 Casimirs
\be
{\hat j} = {{\hat j}_a}^{\pm} {{\hat j}^a}{}^{\pm}= {3{\hbar}^2 \over  
4},  \quad {\hat {\bar H}}^2 - \um{{\hat j}_a}^+ {{\hat j}^a}{}^- 
= {{\hbar}^2 \over 2}.   
\label{27}
\ee

This particular value of the Casimir ${\hat j}$ (in Eq.~\ref{27})
corresponds to a particular discrete representation of $SU(1,1)$ which
will be discussed elsewhere.
Note that the only ordering ambiguity is in ${{\hat j}_2}{}^{\pm} $ 
(Eq.~\ref{25}) but that any other ordering would only produce terms of 
$O(\hbar^2)$ on the R.H.S. of Eqs.~\ref{26} and \ref{27}.
\subsection{The Quantum Modular Group Generated}
The modular group acts classically on the torus modulus and momentum 
and holonomy parameters as
\begin{eqnarray}
&S&: m\rightarrow -m^{-1} ,\quad
     \pi\rightarrow \bar m^2 \pi,
\quad r_1^{\pm}\rightarrow r_2^{\pm},\quad 
    r_2^{\pm}\rightarrow - r_1^{\pm},\nonumber\\
&T&: m\rightarrow m+1 ,\quad \pi\rightarrow \pi ,
\quad r_1^{\pm}\rightarrow r_1^{\pm} + r_2^{\pm},\quad  
    r_2^{\pm}\rightarrow r_2^{\pm},
\label{28}
\end{eqnarray}
and generates the entire group of large diffeomorphisms of 
$\IR\!\times\!T^2$. 

With the ordering of Eq.~\ref{13} (the only ambiguity), the quantum action 
of the modular group is the
same as the classical one, with no $O(\hbar)$ corrections, and is
generated by the $\hbox{SO}(2,2)$  anti-de Sitter subgroup by 
conjugation with the operators $U_T $  and $U_S$ \cite{nr3} where
\bea
U_T &=& \exp {{i \over 2\hbar}(j_0^{\pm} +j_1^{\pm})}= 
\exp {\mp{i\a \over 2\hbar}C_2^{\mp}} 
= \exp {\mp {i\a \over 2\hbar}(r_2^{\pm})^2},\nonumber \\
U_S &=& \exp {{i\pi \over 2\hbar}j_0^{\pm}}= 
\exp {\mp{i\pi\a \over 4\hbar}(C_1^{\mp} + C_2^{\mp})} 
= \exp {\mp {i\pi\a \over 4\hbar}((r_1^{\pm})^2) 
+(r_2^{\pm})^2)}.
\label{29}
\eea

The quantum algebra (Eq.~\ref{26}) and the identities (Eq.~\ref{27})
are invariant under the transformations of Eq.~\ref{28}.
\section{The Extended Algebra}
Using Eq.~\ref{24} it can be checked that the Hamiltonian 
$\hat {\bar H}$ (Eq.~\ref{15}) does not commute with all the 
$\hbox{sl(2,$\IR$)}$ generators (Eq.~\ref{25}), but instead defines 
a new globally constant three-vector ${\hat v}_a$ 
\be
[\hat {\bar H},{\hat j}_a^{\pm}] = \pm i {\hbar} {\hat v}_a, \quad where
\label{31}
\ee
\be
{\hat v}_0=-{\a \over 2}({\hat r}_1^+ {\hat r}_1^- + {\hat r}_2^+ 
{\hat r}_2^-), \quad
{\hat v}_1={\a \over 2}({\hat r}_1^+ {\hat r}_1^- - {\hat r}_2^+ 
{\hat r}_2^-), \quad
{\hat v}_2=-{\a \over 2}({\hat r}_1^+ {\hat r}_2^- + 
{\hat r}_2^+ {\hat r}_1^-).
\label{32}
\ee
The extended algebra of the {\it ten} $\hat {\bar H}, {\hat j}_a^{\pm}, 
{\hat v}_a (Eqs.~\ref{15}, \ref{25}, \ref{32}), a=0,1,2$ then closes 
as follows
\be
[\hat {\bar H},{\hat v}_a] = {i\hbar \over 2}({\hat j}_a^+ - 
{\hat j}_a^-),~
[{\hat v}_a,{\hat v}_b] = -{i \hbar \over 2} \epsilon_{abc}
({{\hat j}^c}{}^+ + {{\hat j}^c}{}^-),~
[{\hat j}_a^{\pm},{\hat v}_b] = i \hbar (\mp  \eta_{ab} \hat {\bar H} 
+ \epsilon_{abc} {\hat v}^c),
\label{33}
\ee
in addition to Eqs.~\ref{26} and \ref{31}, with the 3 additional 
identities (making, with Eq.~\ref{27}, a total of 6 identities)
\be
{\hat v}^a{\hat j}_a^{\pm} = {\hat j}_a^{\mp}{\hat v}^a = 
\pm {3i\hbar \over 2}\hat {\bar H}, \quad 
{\hat v}_a{\hat v}^a ={\hat {\bar H}}^2 - {\hbar^2 \over 2}.
\label{34}
\ee 
The above 10-dimensional algebra is isomorphic to the Lie algebra 
of $\hbox{so}(2,3)$, whose corresponding group is the conformal group of 
3-dimensional Minkowski space \cite{haj}. The precise identifications 
with the dilatation $D$ and the conformal accelerations are given in 
\cite{mn}. Note that, in contrast 
to the generators ${\hat j}_a^+$ and ${\hat j}_a^-$ (Eq.~\ref{25})
of the two commuting $\hbox{sl(2,$\IR$)}$ subalgebras (Eq.~\ref{26}),
the vector ${\hat v}_a$ (Eq.~\ref{32})and $\hat {\bar H}$ (Eq.~\ref{15})
require {\it both } the commuting ${\pm}$ spinors (Eq.~\ref{24}). Their 
action on the ADM and holonomy parameters is under study.
\vskip 0.7truecm
\section*{Acknowledgments}
This work was supported in part  
by INFN Iniziativa Specifica FI41, NSF 
grant PHY-9503133, the European Commission programmes 
TMR-ERBFMRX-CT96-0045, HCM-CHRX-CT93-0362,and the 
Erwin Schrodinger Institute (Vienna).


\section*{References}
\begin{thebibliography}{99}
\bibitem{mn}V. Moncrief and J.E. Nelson, `Constants of motion and 
the conformal anti - de Sitter algebra in (2+1)-Dimensional Gravity', 
gr-qc/9707032, {\em Int. J. Mod. Phys. D} {\bf 6}, 5 (1997).
\bibitem{cn}S. Carlip and J.E. Nelson, {\em Phys. Lett. B} {\bf 324}, 299 
(1994); {\em Phys. Rev. D} {\bf 51} 10, 5643 (1995).
\bibitem{vm1}V. Moncrief, {\em J. Math. Phys.} {\bf 30}(12), 2907 (1989).
\bibitem{hos}A. Hosoya and K. Nakao, {\em Class. Qu. Grav.} {\bf 7}, 163 
(1990); {\em Prog. Theor. Phys.} {\bf 84}, 739 (1990); Y. Fujiwara and 
J. Soda, {\em Prog. Theor. Phys.} {\bf 83}, 733  (1990).
\bibitem{nr1}J.E. Nelson and T. Regge, {\em Nucl. Phys. B} {\bf 328}, 
190 (1989);J.E. Nelson,  T. Regge and F. Zertuche, {\em Nucl. Phys. B}
{\bf 339}, 516 (1990). 
\bibitem{vm2}V. Moncrief, {\em J. Math. Phys.} {\bf 31}(12), 2978 (1990);
\bibitem{mart}S. Martin, {\em Nucl. Phys. B} {\bf 327}, 178 (1989).
\bibitem{car}S. Carlip, {\em Phys. Rev. D} {\bf 42}, 2647 (1990);{\em 
Phys. 
Rev. D} {\bf 45}, 3584 (1992); {\em Phys. Rev. D} {\bf 47}, 4520 (1993).
\bibitem{puz}R. Puzio,  {\em Class. Qu. Grav.} {\bf 11}, 609-620 (1994).
\bibitem{nr3}J.E. Nelson and T. Regge, {\em Phys. Lett. B} {\bf 272}, 
213 (1991);S.Carlip and J.E.Nelson, in preparation.
\bibitem{haj}A related, classical discussion is by P. Hajicek, 
`Group-Theoretical Quantisation of 2+1 Gravity in the Metric-Torus 
Sector' gr-qc/9703030.

\end{thebibliography}
\end{document}